# Bridging the Gap between Technical Heterogeneity of Context-Aware Platforms: Experimenting a Service Based Connectivity between Adaptable Android, WComp and OpenORB

**Valérie Monfort[1, 2,3] and Sihem Cherif[3]**

[1] **Laboratoire SOIE, Tunisia**
*valerie.monfort@univ-paris1.fr*

[2] **Université Paris 1 Panthéon Sorbonne,**
**90 rue de Tolbiac 75634 Paris cedex 13 France**

[3] **Institut Supérieur d'Informatique de Gestion Kairouan, Tunisia**
*cherifsihem16@gmail.com*

**Abstract**

Many companies include in their Information Systems (IS) several communicating heterogeneous middleware according to their technical needs. The need is the same when IS requires using context aware platforms for different aims. Moreover, users may be mobile and want to receive and send services with their PDA that more often supports Android based Human Man Interface. In this paper, we show how we extend Android to make it adaptable and open. We also present how we communicate between different heterogeneous context aware platforms as WComp and OpenORB by using Android and Web Services. We introduce a concrete case study to explain our approach.

## 1. Introduction

Economical context impacts companies and their Information System (IS). Companies acquire other competitors or develop new business skills, delocalize whole or part of their organization. Moreover they are faced to powerful competitors, and they have to shortly develop new products, as less than 3 months, that fit to customer needs. There is faced to these complex evolutions and has to overcame these changes. Service Oriented Architecture (SOA) offers a great flexibility to IS. Each application owns interfaces masking implementation details. Applications are seen as black boxes independently connected to an applicative as Enterprise Application Integration bus (EAI) with its adapters (connecting the bus to the applications).

However, this integration solution does not allow connecting heterogeneous applications or infrastructures, as distant IS. Web services [6] [11] are the cheaper and simplest solution to resolve this problem. They offer interoperability because they are based on standards as XML [2] and allow loose coupling. We proposed aspects based solutions to gain in code simplicity without re deploying code with a non intrusive manner [8] [12].

We based our more recent approach on extended BPEL (Business Process Execution Language) [2] and temporized automatons [1] [7], that we prototyped by providing client, and server adaptability. Moreover, these are also use to manage contextual data coming from different equipments as supervision infrastructure for instance. Current middleware (EAI and ESB) are not fitted to deal with these kinds of information as sending alarms and taking decision. Context adaptation platforms as WCOMP [16], OpenORB [19], Aura [23], Cortex [24], and OpenCom [19]... aim to manage contextual data. On top of all, users are most of the time mobile and they want to access specific and fitted services according to their profile and their location with push and pull manners. Unfortunately, Human Man Interface platforms as Android do not allow adaptability. Concretely, Android [27] is one of the most famous environments used for PDA. The Android platform uses many different technologies.

Some of them are new, and some have been seen before in other settings as: i) Location awareness, through inexpensive GPS devices, ii) Handheld accelerometers, such as those found on the Nintendo Wii remote, iii) Mashups, often combining maps with other information. Several popular Android programs use these concepts to create a more compelling and relevant experience for the user. For example, the Locale application can adapt the settings on the user's phone based on where he is.

Introducing such platforms in IS shows some problems as: i) Interoperability between context aware platforms as WComp and OpenORB, ii) Interoperability with other applications and other middleware, iii) using Human Man Interface that communicates with any platform by messages sending, iv) making current Human Man Interface technologies as Android adaptable according to context. In this research paper we aim to propose an Aspects and Web Services approach to make Android adaptable according to context and to communicate between Android, WComp and OpenORB. This paper is structured as followed. Firstly, we present main technologies we used in our research work. Secondly, we present a case study. Thirdly, we show how we introduced aspects in Android. Fourthly, we propose to add aspects in Android code to increase adaptability. In the Fifth section we show how to bridge the gap between heterogeneous platforms as Android, WComp and OpenORB. Then, we present related works, future works and we conclude. First of all, let us describe now the main technologies we use.

## 2. Basic concepts

### 2.1 Web services

Web services (WS) [10] [11] [15], like any other middleware technologies, aim to provide mechanisms to bridge heterogeneous platforms, allowing data to flow across various programs. The WS technology looks very similar to what most middleware technologies looks like. Consequently, each WS possesses an Interface Definition Language, namely WSDL, which is responsible for the message payload, itself described with the equally famous protocol SOAP, while data structures are explained by XML [19]. Very often, WS are stored in UDDI registry. In fact, the winning card of this technology is not its mechanism but rather the standards upon which it is built. Indeed, each of these standards is not only open to everyone but, since all of them are based on XML, it is pretty easy to implement these standards for most platforms and languages. For this reason, WS are highly interoperable and do not rely on the underlying platform they are built on, unlike many ORPC. According to a vast majority of industrial leaders, WS is the best fitted technology for implementing Service Oriented Architectures.

WSs provide a minimalist mechanism to interconnect different applications. But one fundamental point is the importance of the WSDL being the exact interface of the system. As we said earlier, most of ORPC take a great care of hiding the message layer details from the developer. This approach breaks down when the applications involved do not lay on the same middleware infrastructure, and when interoperability becomes a major concern, traditional ORPC fail to achieve this properly. With WSs, the message contract (WSDL) is the central meeting point which connects applications. The WSDL contract constitutes the design view upon which developers can generate both client and server sides (proxy and stub), as can be seen in Fig. **1**.

WS-BPEL is a WS orchestration language. An orchestration specifies an executable process that involves message exchanges with other systems, such that the message exchange sequences are controlled by the orchestration designer. WS-BPEL provides a language for the specification of Executable and Abstract business processes [10]. By doing so, it extends the WS interaction model and enables it to support business transactions.

Even if WS are the fittest current solution for interoperability in companies IS and SOA implementation, we noticed several limitations. It is the reason why we used Aspect Oriented Programming (AOP) to increase flexibility of SOA.

### 2.2 Aspects

Aspect Oriented Programming (AOP) is viewed as an answer to improve Web services flexibility. AOP [9] is a paradigm that enables the modularization of crosscutting concerns into single units called aspects, which are modular units of crosscutting implementation.

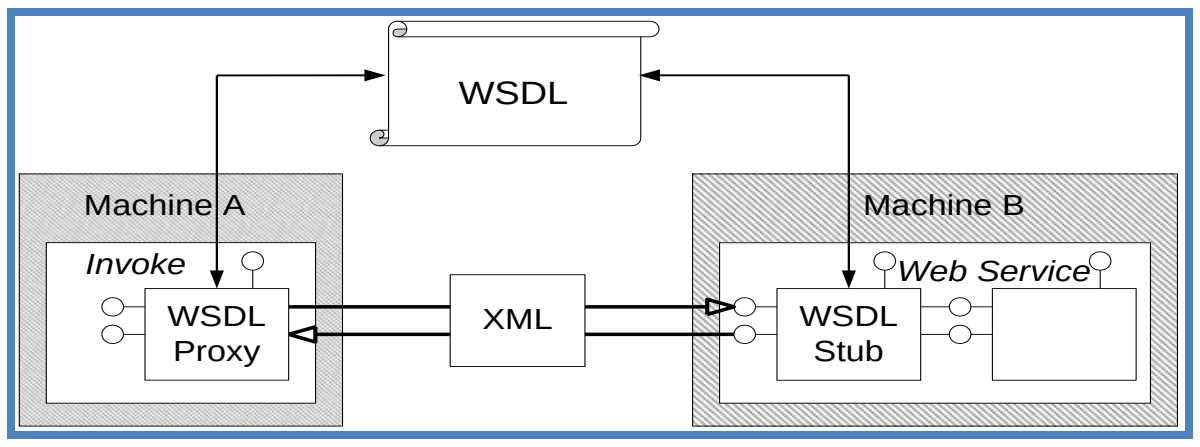

Fig. 1 Web Service Approach.

Aspect-oriented languages are implemented over a set of definitions:
- Joinpoints: They denote the locations in the program that are affected by a particular crosscutting concern.
- Pointcuts: They specify a collection of conditional joinpoints.
- Advices: They are codes that are executed before, after or around a joinpoint.

In AOP, a tool named weaver takes the code specified in a traditional (base) programming language, and the additional code specified in an aspect language, and merges the two together in order to generate the final behaviour. The weaving can occur at compile time (modifying the compiler), load time (modifying the class loader) or runtime (modifying the interpreter). Connecting applications requires some specific technologies as adapters, as presented as followed.

## 2.3 Adaptation and plasticity

Adaptation is modeled as two complementary system properties: adaptability and adaptivity. Adaptability is the capacity of the system to allow users to customize their system from a predefined set of parameters. Adaptivity is the capacity of the system to perform adaptation automatically without deliberate action from the user's part. Whether adaptation is performed on human requests or automatically, the design space for adaptation includes three additional orthogonal axes.

-The target for adaptation: This axis denotes the entities for which adaptation is intended: adaptation to users, adaptation to the environment, and adaptation to the physical characteristics of the system. The physical characteristics of a system can be refined in terms of interactional devices (e.g., mouse, keyboard, screen, video cameras), computational facilities (e.g., memory and processing power), and communicational facilities (e.g., bandwidth rate of the communication channels with other computing facilities).

-The means of adaptation: This axis denotes the components of the system involved in adaptation: typically, the system task model, the rendering techniques, and the help subsystems, may be modified to adapt to the targeted entities. The system task model is the system implementation of the user task model specified by human factor specialists. The rendering techniques denote the observable presentation and behavior of the system. The help subsystems include help about the system and help about the task domain.

-The temporal dimension of adaptation: Adaptation may be static (i.e., effective between sessions) and/or dynamic (i.e., occurring at run time).

The term "plasticity" is inspired from the property of materials that expand and contract under natural constraints without breaking, thus preserving continuous usage. By analogy, plasticity is the capacity of a user interface to withstand variations of both the system physical characteristics and the environment while preserving usability. In addition, a plastic user interface is specified once to serve multiple sources of variations, thus minimizing development and maintenance costs. Plasticity may be static and/or dynamic. It may be achieved automatically and/or manually. Within the design space of adaptation, plasticity is characterized in the following way:

- Along the target axis, plasticity is concerned with the variations of the system physical characteristics and/or the environment. It does not, therefore, cover adaptation to users' variations;
- Along the means axis, plasticity involves the modification of the system task model and/or of the rendering techniques;
- The temporal and automaticity axis are left opened.

Technically, plasticity requires software portability. However, platform independent code execution is not a sufficient condition. Virtual toolkits, such as the Java abstract machine, offer very limited mechanisms for the automatic reconfiguration of a user interface in response to variations of interactional devices. All of the current tools for developing user interfaces embed an implicit model of a single class of target computers (typically, a keyboard, a mouse and at least a 640x480 color screen).

As a result, the rendering and responsiveness of a Java applet may be satisfactory on the developer’s workstation but not necessarily usable for a remote Internet user. Experience shows that portability does not guarantee usability continuity. In addition, the iterative nature of the user interface development process, as well as code maintenance, makes it difficult to maintain consistency between the multiple target versions.

The user profile [13] covers broad aspects such as its cognitive and social environment, which determine its intentions during a session of research [12]. The construction of the profile reflects a process that allows to instantiate its representation from various sources. This process, generally implicit, is based on an inference process on the user’s context and preferences via his behavior during the use of: i) An access system to information requests, ii) A Web browser, iii) Other e-mail applications tools. The evolution of the profiles [14] indicates their adaptation to the variation of the users’ interests they describe, and consequently, the variation of their services and information needs in time. Let us explain now what Android is.

## 2.4 Context Aware Platforms

The term context is defined as any information that can be used to characterize the situation of an entity. An entity is a person, place or object that is considered relevant to the interaction between a user and an application, including the user and application themselves [28]. Some examples of context include location, company policy, resource availability, hardware and software environment, physical environment, user identity and the goals of the user.

OpenORB [19] is an adaptable and reflexive middleware. OpenORB provides a Java implementation of the OMG CORBA 2.4.2 specification including following services as: Concurrency Control Service, Event Service, Interoperable Naming Service, Notification Service, Persistent State Service, Property Service, Time Service, Trading Service, and Transaction Service. OpenORB has been designed to provide a reliable foundation for distributed applications. It combines all CORBA features with implementation specific extensions, with the aim of being the most powerful and complete CORBA implementation in Java. OpenORB is the successor of JavaORB that is already widely used all over the world: in Europe, America, China, Australia, and other places. A large number of deployed applications, research projects, and study projects are using JavaORB. Building on their experience with JavaORB, the OpenORB team has defined a complete new architecture to ensure that OpenORB is the best solution for applications needing high scalability and high performance.

In addition, WComp [16] is a prototyping "development" environment for context-aware applications. The WComp Architecture is organized around Containers and Designers paradigms. The purpose of the Containers is to take into account system services required by Components of an assembly during runtime: instantiation, destruction of software Components and Connectors. The purpose of the Designers allows configuring assemblies of through Containers. To promote adaptation to context WComp uses Aspect Assembly paradigm. Aspect Assemblies can either be selected by a user or fired by a context adaptation process. It uses a weaver that allows adding and or suppressing components. A container includes a set of (Beans) components and each bean has: properties, input methods that use received input information, and output Methods to send to another bean, for instance, output information. Aspect Assemblies allow defining links between Beans by using input and output information. WComp uses UPnP (Plug and Play) technology to detect locally whether the device is active or not and to define input methods and sent events for each component. With this architecture WComp allows: i) managing devices heterogeneity and dynamic discovering by using UPnP, ii) events driven interactions with devices, iii) managing dynamic devices connection and disconnection (dynamic re configuration on run time) in infrastructure. Let us see now the proposed solution. We notice these platforms are heterogeneous, because one is based on reflexivity and .jar configuration file (Open ORB) and the other one is based on dynamic assembling and .dll configuration file (Wcomp) as shown in array 1.

Table 1. Main characteristics of the context adaptation platforms

| Criteria | | WComp | OpenORB |
|---|---|---|---|
| Architecture | Centralized | X | |
| | Decentralized | | X |
| Service | | X | |
| Events management | | X | |
| Object/Componen t | Object | | X |
| | Component | X | |
| Re configuration | Reflexivity | | X |
| | Assembling | X | |
| Configuration file | .dll | X | |
| | .jar | | X |
| Interoperability | | X | X |
| Adaptability approach | MOP | | X |
| | Weaving | X | |
| | Generation | X | |

To illustrate our approach, let us present now our case study.

## 3. Case Study

### 1. Case Study Description

A person owns a PDA and walks in the street. This person subscribed to different service provider's access to different services dynamically provided according to location but also according to the will of the person. Some other services in a SaaS [29] manner are also offered to the person according to his profile: sex, age, profession, handicapped or not... Moreover, the person can precise the distance between him and the services. For instance, the person subscribed to bank services. He walks and the vocal system asks "Do you want to find a bank", the person can answer "yes". The system can propose the closest bank and indicates the way to go to the bank. The person can ask for another bank one kilometer around. The system proposes to him other banks and asks for him what kind of operations he wants to do. The person can ask for an ATM, and the system can indicate the ATM of the selected bank does not work; the person has to choose another proposed bank. Moreover, the HMI adapts itself according to the profile of the person. For instance, for a girl the Interface will be in pink, for a boy in blue. If the person is blind, the vocal interface will be switched on. The person can ask to dynamically change the color or the display mode. This person may use his PDA to manage the different devices in his smart house and he can switch on the light, TV ... remotely. Moreover, this person works in a company that manages different devices and he has to administrate the equipments. The equipments send information to a middleware that sends information to a database or if there is a problem the user receives a message on his PDA else on his TV screen if the PDA is switched off.

## 2. Proposed Architecture

WComp is the middleware found in the Smart House where services are embedded. OpenORB is used in the company as a middleware to collect information and alarms from the equipments to send them in a data base and/or on the PDA of the user. WComp, OpenORB and Android have to communicate even if they are not based on the same technologies. We use Sharp-Develop software to implement the case study. So, we can solve the problem of interoperability between platforms, and we can connect the three platforms Open ORB, Android and Wcomp using Web services. The company has a gateway and the enterprise portal is one of the bricks of the information system of the companies as each customer reaches this gate with his account. The notion of portal became important when the customer is mobile, thus, it is going to reach in the portals of companies via his PDA. So, the main process is as followed: i) To adapt the Human Man Interface according to location and profile, ii) To show the alarms via OpenORB, iii) To recover the messages of alarms via user's PDA, iv) To send the alarms from PDA via WCOMP in the Smart House, iv) If PDA is off OpenORB sends the messages on the TV screen. Communication and interoperability is allowed by Web Services (Fig. 2).

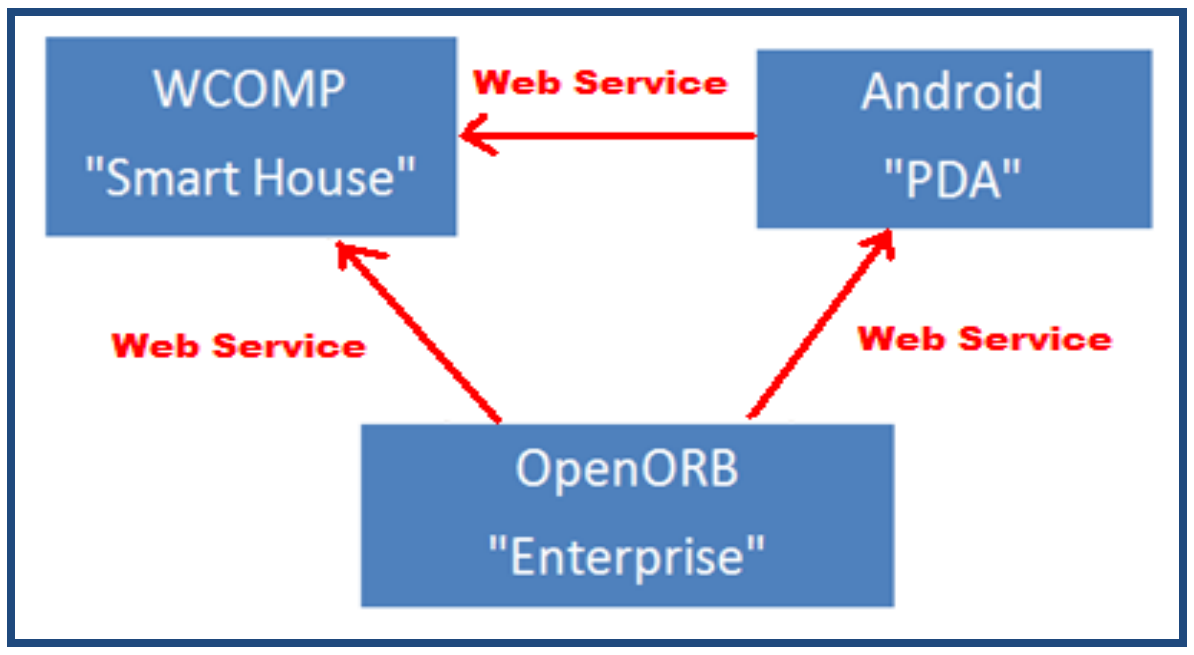

Fig. 2 General architecture

Android side architecture (Fig.3) proposes to embed Android on the PDA. Android dynamically adapts presentation, it allows asking for services, it allows sending location and user ID (step 1) to an identification gateway (step 2) that identifies and authenticates the user to know his profile (step 3). Then, profile information are sent to Android that adapt the HMI(step 4) and the system looks for the fitted available services (step 5) that are sent to the user via Android (step 6). So services are sent on a pulled or pushed manner

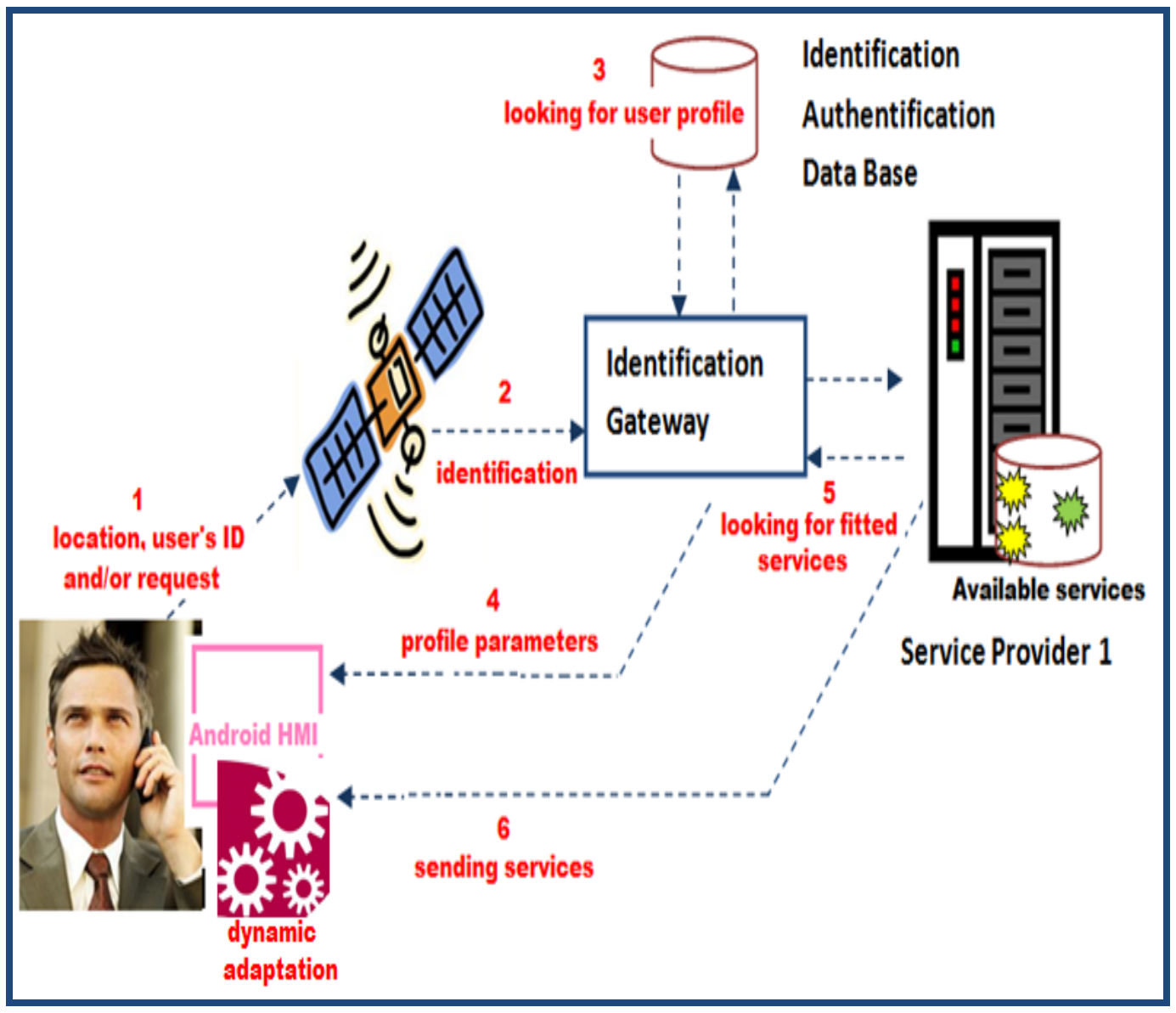

Fig. 3 Android side architecture

Fig.4 shows the different equipments of the company sending messages to the OpenORB middleware in the Information System (IS) of the company (step1). It collects messages and according to the status of the message, it sends critical messages to the user on his PDA (step 2) via Android platform. The user may be mobile as walking in the street or be in his house, so the alarms are sent to his PDA in the Smart House (step 3). If the PDA is off, alarms are sent from OpenORB via Wcomp that sends the alarms via TV screen if it is switched on.

# 4. Android Adaptability Prototype

## 4.1. Android project files

Each Android project includes three main files as: i) Main.XML that manages interface components, ii) AndroidManifest.XML that contains the set of references used to execute Android application, iii) Active.java that is used to start Android.

## 4.2. Android project creation

As proposed in Code 1, the name of the project is "Android_Location_Profil_Services" with the characteristics of the created project.

```
Project name:   Android_Location_Profile_Service
Build Target:   Android 1.6
Application name:   Android_Location_Profile_Service
Package name:   com.Android_Location_Profile_Service
Create Activity:   Android_Profile_Service
```

Code 1. Characteristics of the Android created project

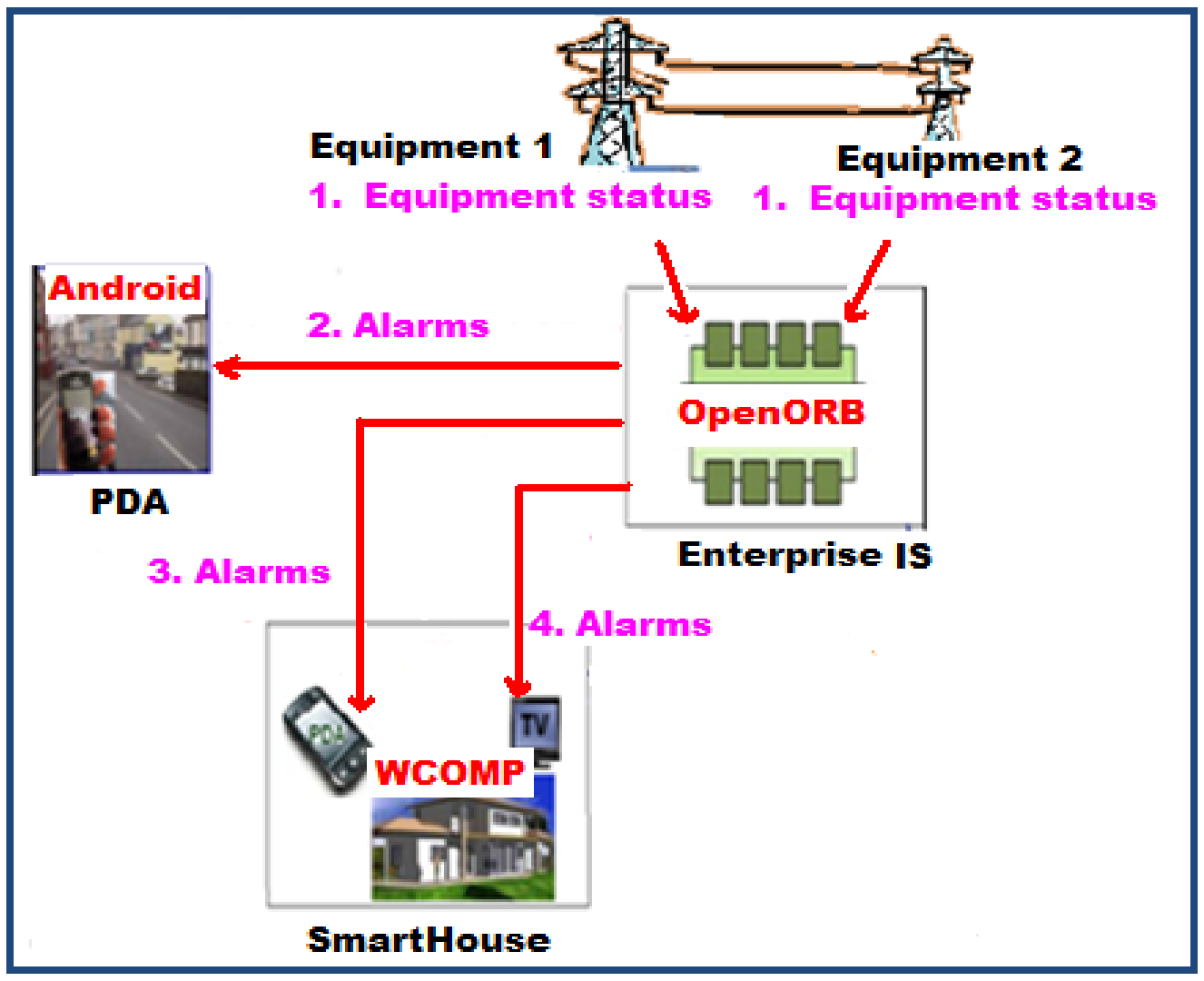

Fig. 4 Communications between platforms

## 4.3. Creation of the "Before" aspect

Before the method "onCreate" (Code 2) an aspect is created that displays the following message: "Provide available services please:".Line 2 shows the aspect "BeforeService". Line 3 contains a "pointcut" that intercepts the method of the class "Android_Profil_Service" from "com. Android_Location_Profil_Service" package.

```
1. package
   com.Android_Location_Profile_Service ;

2. aspect BeforeService {

3. pointcut aspbefore ():execution
   (*com.Android_Location_Profile_Service.
   Android_Profile_Service. onCreate (..));

4. before () : aspbefore (){

System.out.println ("**********Provide
available services please : ***********");
```

Code 2. Aspect creation

The extension of the Spring IDE plug-in includes Spring IDE AOP Extension. Spring IDE AOP offers the Bean Cross References view that allows watching the various aspects weaved on one bean, and, on which bean, an aspect is weaved. This view owns a hierarchical structure to display all the various aspects as well as the contained code advice. Then, for each of them, is attached the information linked to beans weaved with the impacted methods. On the other hand, the same hierarchical structure offers the possibility to display, for one selected bean, the lists of the aspects weaved on these methods with an indication on the type of the corresponding code advice. In fine, Cross References allows us giving the list of methods with them aspects (Fig.5). The "Cross References" makes sure that the module Spring AOP is well to start.

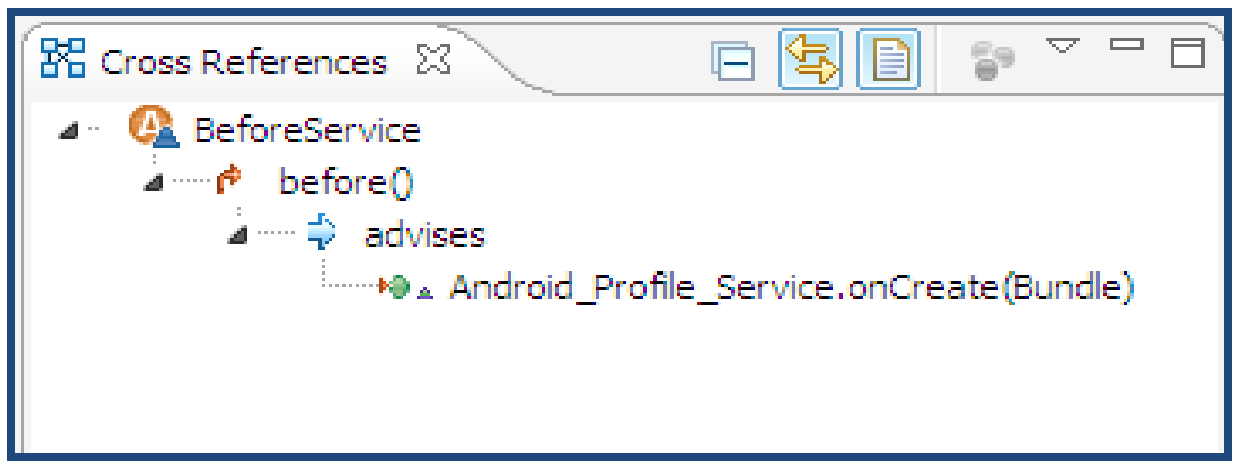

Fig. 5 Cross references

## 4.4. Modification of "AndroidManifest.xml"

The access to certain critical operations is restricted, and it is necessary to specifically ask for permission to use them in a file named called Android- Manifest.xml. When the application is installed, the Package Manager either grants or doesn't grant the permissions based on certificates and, if necessary, user prompts. Here are some of the most common required permissions:

- INTERNET: Access the Internet.
- ACCESS_COARSE_LOCATION: Use a coarse location provider such as cell towers or WIFI.
- ACCESS_FINE_LOCATION: Use a more accurate location provider such as GPS.

Using XML tags in AndroidManifest.xml allows restricting who can start an activity, start or bind to a service, broadcast intents to a receiver, or access the data in a content provider. So, we add these lines to "AndroidManifest.xml" before the XML tag <application>. The system uses this permission before running the application. Lines 1 to 3 (Code 3) allows an application accessing to a secondary location (as WIFI for instance). Lines 4 to 6 allow an application to access the fitted site (as GPS). Lines 7 and 8 allow an application accessing to Internet.

```
<uses-permission
android:name="android.permission.ACCESS_COARSE_LOCATION">
 </usespermission>
<uses-permission
android:name="android.permission.ACCESS_FINE_LOCATION">
 </usespermission>
<uses-permission android:name="android.permission.INTERNET">
</usespermission>
```

Code 3. AndroidManifest.xml

## 4.5. Modification of "main.xml"

« main.xml » is an Android XML file. Android is optimized for mobile devices with limited memory and horsepower, so you may find it strange that it uses XML so pervasively. But XML is considered as a verbose, human-readable format not known for its brevity or efficiency. The Eclipse plug-in invokes the Android resource compiler, to adapt, and to preprocess the XML into a compressed binary format. It is this format, not the original XML text that is stored on the device. Line 1 (code 4) offers to create a component "TextView" called "Location". From this component, you can display on screen the text "Location". Line 2 proposes a component "EditText" in the Android interface. The text area "Edit text" uses to fill up the coordinates to select users.

Android uses XML files for the Layout of widgets. In our example, the Android for Eclipse plug-in generates a main.xml file for the layout. This file owns the XML based definitions of the different widgets and their containers.

## 4.6. Creation of the classes: "profile", "service"," location"

We aim creating three data bases as: profile.db, service.db and location.db. Profile.d includes a "profile" table that shows five attributes as: id-profile, name, sex, job, and age. Service.db contains a "service" table that includes five other attributes as: id-service, service-name, description, location and distance. Location.db includes a "location" table that includes three attributes as: id-location, longitude, altitude. We add in the bar of title of the application the identifier of the person during loading of the application (Fig.6).

```
1.  <TextView android:textColor="#000000"
    android:layout_width="wrap_content"
    android:id="@+id/position" android:text="Position"
    android:layout_height="wrap_content">
    </TextView>

2.  <EditText android:layout_width="100dip"
    android:text=" "
    android:id="@+id/editposition"
    android:gravity="center"
    android:layout_height="50dip"
    android:layout_marginLeft="60dip" >

3.  </EditText>

4.  <EditText android:layout_width="100dip"
    android:text=" "
    android:id="@+id/editposition1"
    android:gravity="center"
    android:layout_height="50dip"
    android:layout_marginLeft="10dip" >

5.  </EditText>
```

Code 4. Main.xml

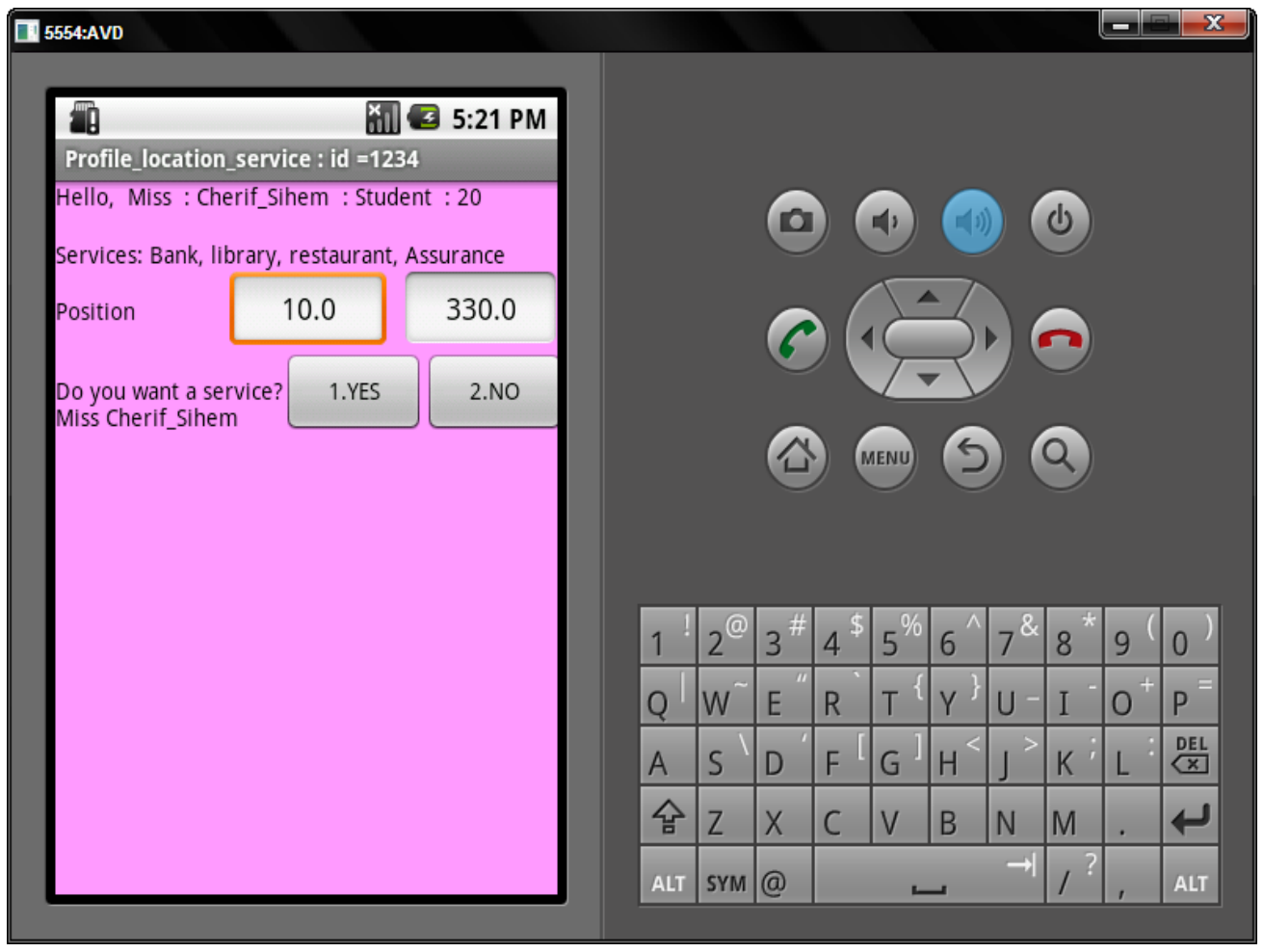

Fig. 6 General interface

## 4.7. Creation of the first interface

From Code 5, we get information concerning the user as name, sex, age ... and the system joins this information to a message called "Hello". The system recovers services (Code 6) saved by the person "Cherif Sihem" in the database according to the profile as shown in Code 7. This method adds to the database a new person. The message "Do you want a service?" is listen by the user who can answer. We aim to get data from profile.db database and to use them in the Android application. For instance, we insert in the text id-profile of the Android activity the id-profile of the profile.db database.

```
String id_profile= cursor.getString(1);
String Name= cursor.getString(2);
String Sexe = cursor.getString(3);

builder.append(Sexe).append(" : ");
builder.append(Name).append(" : ");
builder.append(Age).append("\n ");

// Display on the screen

setTitle( "Profile_location_service: id
="+id_profile);
bonjour.setText("Hello, "+builder);
```

Code 5. Interface building

```
TextView service_by_profile= (TextView)
findViewById(R.id.Services_by_profile);
service_by_profile.setText("Services: Bank,
library, restaurant, Assurance");
```

Code 6. Profile linked to service

```
profil = new EventsData(this);
try {
addEvent("1234","Cherif_Sihem","Miss",
"Student","20");

Cursor cursor = getEventsprofil();
showEventsprofil(cursor);
```

Code 7. Identification of the user and profiling

## 4.8. Results example

The user is detected with a GPS system. The services are displayed according to his profile and his location. The system detects Sihem is a girl without any handicap but she asks for vocal assistance. So, it proposes her a list of services on her screen. To get the list, she presses on the button “1.YES” then a list of services is posted. We only displayed the left part of the PDA screen.

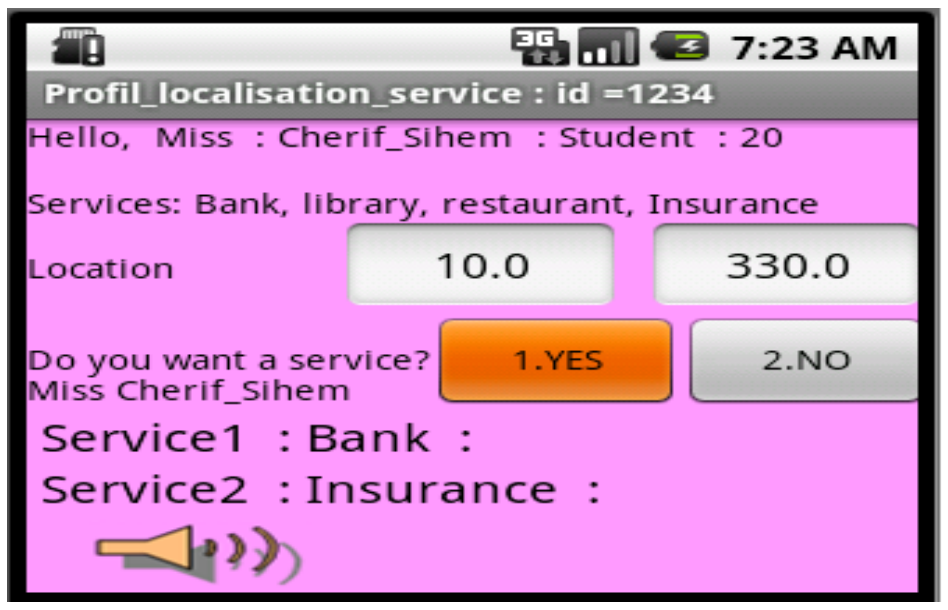

Fig. 7 Proposed services according to location

She says “Insurance” and the system proposes her several possibilities (Fig. 8) as: BIAT, STB and BNA with the distance according to her location. She decided also to change the color of the screen. She chooses BIAT Company.

After having presented the introduction of aspects in Android code to get adaptability, let us propose now to illustrate interoperability between platforms.

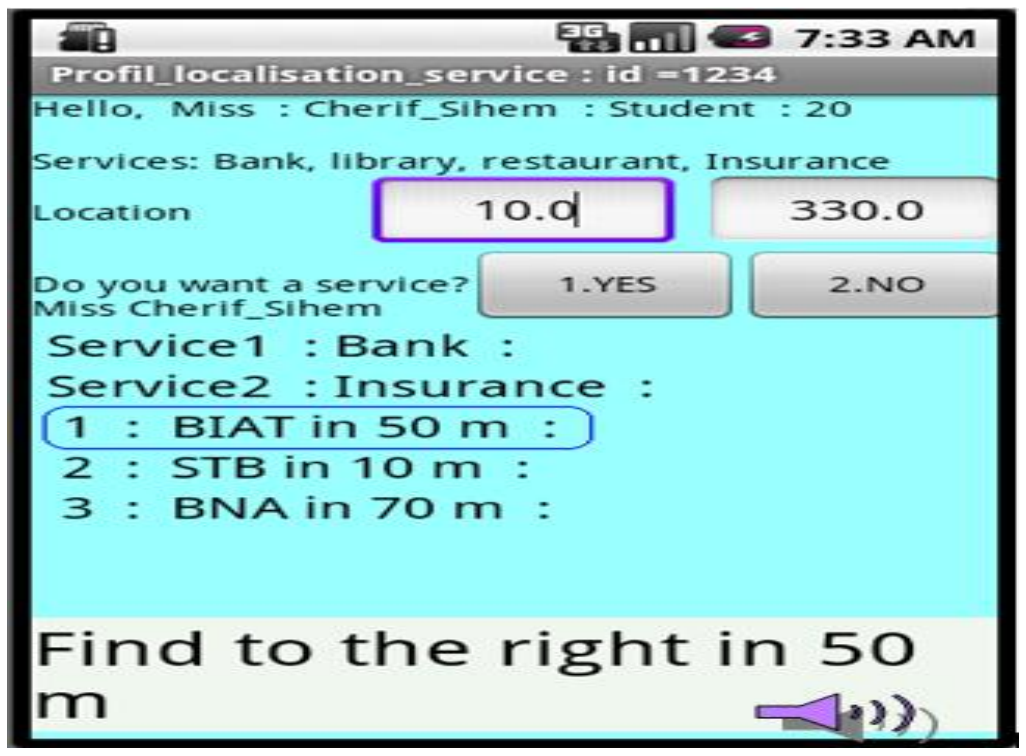

Fig. 8 Proposing services with the distance

# 5. Interoperability Implementation

## 5.1. Parameterization:

This initialization of the system requires several steps shown as followed. To call the services Web on Android, we use KSOAP2 (code8) (ksoap2-android-ensemble-2.4-jar-with-dependencies.jar). Then we start Apache Axis2 server. So our application is executed in Web mode. The ksoap2-android project provides a lightweight and efficient SOAP library for the Android platform. So, the user calls a Web service from Android using KSOAP2

```
import org.ksoap2.SoapEnvelope;
import org.ksoap2.serialization.SoapObject;
import org.ksoap2.serialization.SoapSerializationEnvelope;
import org.ksoap2.transport.HttpTransportSE;
```

Code 8. Using KSOAP2

Each OpenORB project is based on java classes as: server.java, client.java, and on a file. Then, the goal of the OpenORB server is to provide objects (corresponding to the implementation) to the customers, to receive the requests. This operation proceeds in 3 stages: i) Initialization of the ORB, ii) Activation of the POA (the POA - Portable Object Adapter - is an object adapter that allows managing several CORBA objects), iii) End of the execution of the ORB.

Then the creation of the OpenORB client comes. The goal of the customer is to reach the distant object and to call upon the methods suggested by this object. Its operation proceeds in 3 steps: i) Initialization of the ORB, ii)

Invocate the distant method, iii) Obtaining a generic reference towards the distant object (starting from the IOR). Then, we create a basic method to display an alarms generator simulator (alarms normally coming from the equipments). We aim now to transform OpenORB classes into Web Services. Then we shall call these services in an Android project. For instance, we call the service “AfficherNormal”.

Then we create a method to display alarms called Afficher.wsdl (Code9). We create Web services, by using as source code the Java classes of the "Enterprise" application we created before, as shown from line 1 to 3.

Then we modify the activity of Android “Portal” (Code 10). Line 1 allows using the dynamic project Web “Ent” that contains Web services. Line 2, allows defining the Web services method. Line 3, allows declaring an action SOAP. Line 4, of code 7, defines the relative path of the Web service, to do so, it is necessary to use IP address “192.168.1.2” with port 8080 instead of local host.

```
1. <wsdl:operation name="AfficherNormal">

2. < wsdl:input message="ns:AfficherNormalRequest"
wsaw:Action="urn:AfficherNormal" />


3. <wsdl:output message="ns:AfficherNormalResponse"
wsaw:Action="urn:AfficherNormalResponse" />
```

Code 9. Method Afficher.wsdl

```
1. private String NAMESPACE = "http://Ent";
2. private   String METHOD_NAME = "AfficherNormal";

3. private  String SOAP_ACTION = NAMESPACE + METHOD_NAME;
// NAMESPACE + method name

4. private static final    String URL =
"http://192.168.1.2:8080/Entreprise/services/EntImpl?wsdl";
```

Code 10. Activity modification of Android “Portal”

Then, Change “AndroidManifest.xml” We add the line 1, (Code 10), "AndroidManifest.xml" before the tag XML < application >: that allows an application to get Internet.

```
1.     <uses-permission
android:name="android.permission.INTERNET">
       </usespermission>
```

Code 11. Change “AndroidManifest.xml”

When the tag <uses….INTENET> is added it is possible to access to the Android application via Internet and Web Services.

## 5.2. Results: getting connection with enterprise portal and Android to receive alarms

Figure 9 represents the user interface that allows reaching the portal and controlling the status of the equipments. With “Show Message” button, the user can retrieve the messages coming from the enterprise middleware with the Web Services via Android.

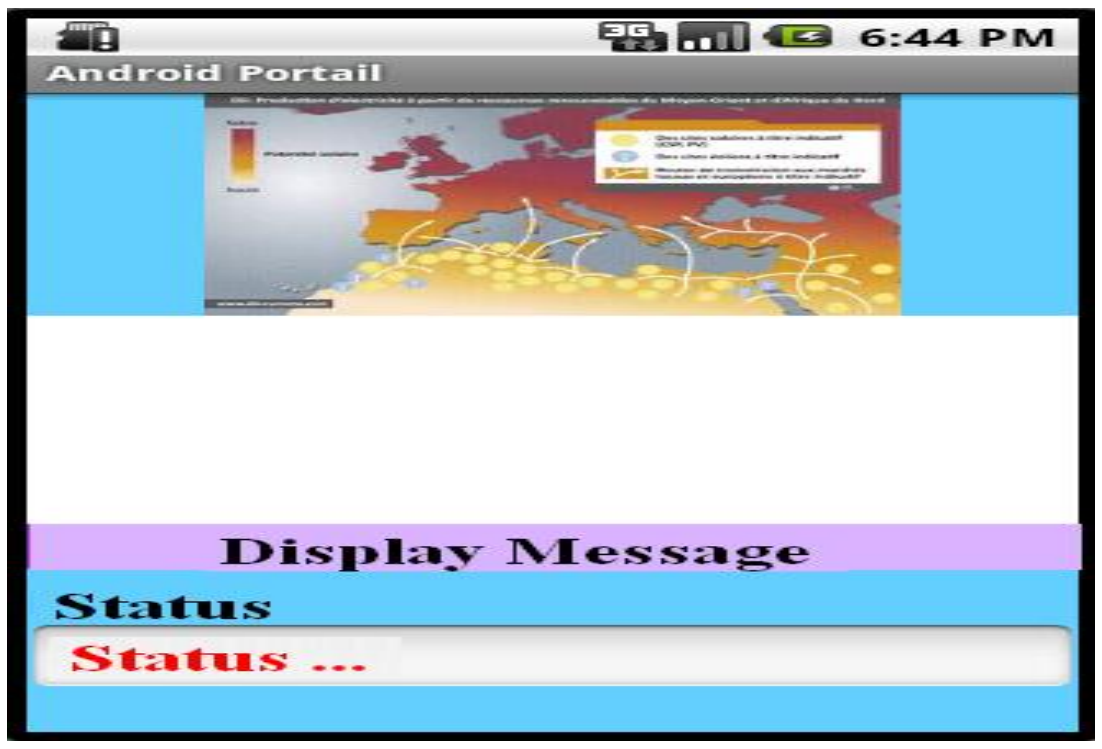

Fig. 9 Android Interface showing company portal and alarms

## 5.3. WComp and OpenORB Connection

The Wcomp modeling tool allows us showing the set of components. We show through a human machine interface, how a user can receive messages from a company from a TV or a PDA. Wcomp in any application is as an assembly of components encapsulated in a composite service. Line 1 (Code 12) invokes a method "AfficherNormal" Web

```
1. public AfficherNormalResponse AfficherNormal() {

2. object[] results = this.Invoke("AfficherNormal", new object[0]);

3. att_AfficherNormal = ((AfficherNormalResponse)(results[0]));

4. this.FireAfficherNormalMethodEvent(att_AfficherNormal);

5. return ((AfficherNormalResponse)(results[0]));
```

Code 12. Displaying service description

Project "Enterprise". Line 2 allows attributing the "results" of the method "AfficherNormal. We invoke OpenORB methods with Web services. With SharpDevelop, we create a new "web service proxy» that admits parameters such as the location of Web Service "WSDL URL" and the name of the workspace. Line 2 describes the way WSDL.

Let us see now the simulation of alarms coming in the Smart House, either on the PDA either on the TV Screen. The process is design with the WComp design tool (Fig. 10). The link (1) with the user interface allows activating “button 1” (assembly) and the component "entreprise31" defining the entry methods the events management associated to components. It allows getting back the OpenORB project. If radio Button called " radiobutton 1 " has the status " check ", then the PDA is switched on and the message will be shown on the PDA, and if radio Button " radiobutton 1 " is in a state " not check ", then the PDA is switched off and the message will be shown on the TV. The link (2) represents the assembly between the component "Enterprise" and the component "EventToggler". The component "EventToggler " allows changing the status of component "radiobutton 1 " to change the status of " check". The link (4) represents the assembly enter the component "radiobutton 1 " and the component " PDA ". The component PDA allows feigning a PDA of an intelligent house which allows showing messages received by the company. At this level, the Web services intervene and allow specifying the event to be produced and to be shown in Textbox «PDA ". The « EventToggler » component allows changing the status of the « PDA » and « TV » components with its specific evenemential methods. I mean to change the status “checked” into “not checked” status and vice versa (link3). The component « SOAP proxy » is a Web Service. The component “PDA” is a textbox showing the PDA interface. The “TV” component is textbox representing the TV interface.

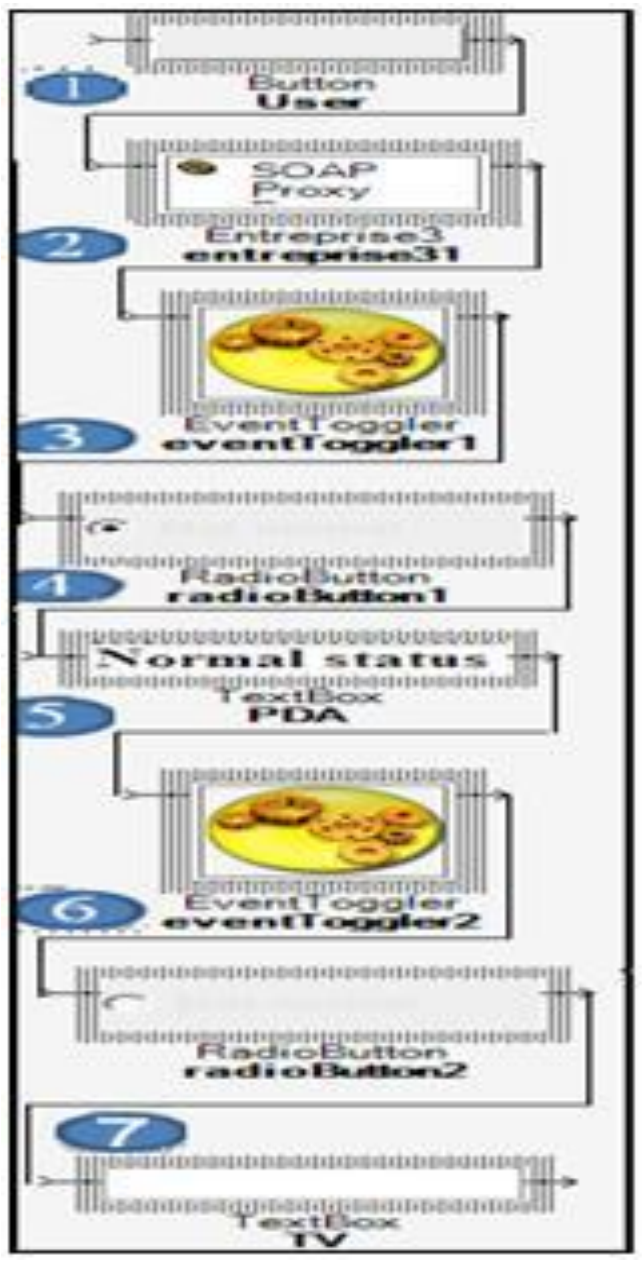

Fig.10 WComp modeling alarms are routed via TV screen

## 6. Related works

Many research works aim to resolve contextual adaptability with different approach as platforms [15], model engineering, ... and HMI are also impacted by these works as shown in [16] [4] and [5], with actors dedicated to plasticity management and taking into account a limited range of contexts, focusing itself on characteristics as the size of the screen, the language… The problem found in the approaches is the difficulty in capturing and processing the contextual data. Mostly, research works aim at handling the user interface without being concerned with functional part.

In addition, several research works aims interoperability management concerning Wcomp, OpenORB and Android:

- [30] introduces the WComp middleware approach, which federates three main paradigms: an event-based Web services approach, a lightweight component-based approach to design composite Web services, and an adaptation approach using the original concept called AA (Aspect Assembly).
- [31] demonstrates that SOA (Service oriented Architecture) and its numerous principles are well adapted for pervasive computing.
- [32] proposes an approach for activity recognition that combines the use of video cameras with environmental sensors to determine as many activities as possible. This approach consists in analyzing human behaviors and looking for changes in their activities. In particular, the goal is to collect and combine multi-sensor information to detect activities. The SUP platform gathers a set of modules devoted to design applications in the domain of activity recognition. Wcomp aims at assembling services which evolve in a dynamic and heterogeneous environment.
- In [33], [34] Ethylene based on Wcomp interests in the software nature of the plastic interactive systems within the framework of the ambient computing. This approach supports the original combination of the approaches with components and with services (aspects reconfiguration and dynamic availability) and the principles of the aspects "interoperability".

Concerning OpenORB, [35] introduced Pontifex, a bridge generator for the connection of CORBA-based applications to Web Services. To reduce the development effort while simultaneously maintaining the quality of the IDL parser, authors decided to rely on OpenORB’s freely available parser as a basis. And [36] combined the reflective component model and the CFbased structuring principle, adding an additional plug-in component type to the protocol CF which enables the configuration of a range of demultiplexing strategies.

Concerning Android, there are very few Android applications to integrate Information System (IS) and business applications. But it seems possible with Web services as proposed in documentations.

## 7. Conclusions

Mobility coupled with the development of a wide variety of access devices has engendered new requirements for HMI such as the ability to adapt to different contexts of use. Moreover, with the apparition of attractive and flexible platforms as Android, HMI development offers a new opportunity to propose to the users multi functions and services including business, M-learning and usual life. Even if we proved Android adaptability and platforms interoperability this approach presents some limitations as it remains a prototype with test cases. We would like to develop a concrete system with context adaption plate forms and data bases in SaaS environment. We also would like to measure performances of such approach and improve security. Moreover, a full context aware approach requires interoperability between platforms to support adaptability. Even if Web Services are the fitted solution to support interoperability nevertheless it remains the challenge of today companies.

**Valérie Monfort** Is an Assistant Professor in Paris 1 Panthéon Sorbonne, working also in Tunisia where she created a research team. She teaches: Information System changes via performance, Mobile Information Systems, SOA based distributed architectures, project management, methodology… She wrote several books in French to disseminate Web service technology. She also wrote several papers about Web service and adaptability. She worked more than twenty years in the industry as an international consultant working for big companies as: IBM, Airbus, DEXIA, BNP, Carrefour, MMA…

**Sihem Cherif**. Is a student master of Science in Intelligent Information system in Kairouan Tunisia. She is starting a career as a scientist and has already published in several famous conferences.